\documentclass[aps,prb, twocolumn,groupedaddress,showkeys]{revtex4-1}
\usepackage{amsmath}
\usepackage{physics}
\usepackage{lipsum}
\usepackage{braket}\usepackage{bm}
\usepackage[pdftex]{graphicx}

\usepackage[colorlinks]{hyperref}
\hypersetup{colorlinks,
citecolor=blue,
linkcolor=blue,
urlcolor=blue,
}

\begin{document}
\title{Time Quantified Monte Carlo Method for Long-range Interacting Systems}
\author{Taichi Hinokihara$^{1,2}$}
% \email{hinokihara@spin.phys.s.u-tokyo.ac.jp}
\author{Yuta Okuyama$^{3}$}
\author{Munetaka Sasaki$^{4}$}
\author{Seiji Miyashita$^{1,2}$}
%\email{miya@spin.phys.s.u-tokyo.ac.jp}
\affiliation{%
	$^1$Department of Physics, Graduate School of Science,
	The University of Tokyo, 7-3-1 Hongo, Bunkyo-Ku, Tokyo 113-8656, Japan\\
	$^2$Elements Strategy Initiative Center for Magnetic Materials(ESICMM), National Institute for Materials Science, Tsukuba, Ibaraki, Japan\\
	$^3$Department of Applied Physics, Tohoku University, 6-6-05 Aoba, Aramaki, Aoba-ku, Sendai 980-8579, Japan\\
	$^4$Faculty of Engineering, Kanagawa University, 3-27-1 Rokkakubashi, Kanagawa-ku, Yokohama 221-8686, Japan\\
}
\begin{abstract}
We propose a method for simulating the stochastic dynamics of classical spin systems with long-range interactions.
The method incorporates the stochastic cutoff (SCO) method, which is originally specialized for simulating equilibrium state, into time quantified Monte Carlo (TQMC) method.
We analytically prove that the present method gives the same real-time dynamics with the stochastic Landau-Lifshitz-Gilbert (s-LLG) equation, i.e., both method derives the same Fokker-Planck coefficients.
We demonstrate magnetization reversal processes and confirm that the result is in good agreement with the result obtained by s-LLG.
Using our method enables us to analyze complicated lattice systems consisting of many spins in a unit cell.
Technical improvement of TQMC is also proposed.
\end{abstract}
\keywords{dynamics of classical spins; long-range interacting system; time-quantified Monte Carlo; stochastic cutoff method}

\maketitle
\section{Introduction}
\label{sec:intro}

Recently, numerical simulations based on atomistic-scale systems are widely developing thanks to the development of the first-principles calculations.
While simulations on the atomistic models give us specific properties of materials, computational cost is increased due to a complicated lattice structure in which large number of atoms in the unit cell $N_{\mathrm{atom}}$ are contained.
In particular, this problem becomes serious when the dipole-dipole interaction (DDI) plays an important role, e.g., evaluating a coercivity of magnet~\cite{Toga2016,Nishino2017,Toga2018} and analyzing the muon spin rotation/relaxation ($\mu$SR)~\cite{Hayano1979,Blundell1999,Moller2013,Bonfa2016,Suprayoga2018}.
For these purposes, numerical methods that can treat DDI in complicated lattice structures are required.

Thus far, the long-range interacting spin system is mainly studied in two approaches: equilibrium state calculation using Monte Carlo (MC) method and stochastic dynamics based on the Langevin equation (stochastic Landau-Lifshitz-Gilbert, s-LLG).
For the case of equilibrium state calculations, several efficient methods have been proposed~\cite{Mak2005a,Sasaki2008,Fukui2009,Sasaki2010,Endo2015a,Hinokihara2018}.
Among them, in our previous work, we proposed a new algorithm for the stochastic cut-off (SCO) method, which can calculate complicated lattice systems efficiently~\cite{Hinokihara2018}.
On the other hand, for the spin dynamics simulations, using the fast Fourier transformation (FFT) is the only efficient way of calculating the long-range interacting systems without any approximations~\cite{Inami2014,Tsukahara2017}.
However, as the simulation time advances, the convolution integrals in FFT need to be run $N_{\mathrm{atom}}^2$ times.
This fact makes us difficult to simulate complicated lattice systems using FFT.

In this situation, it could be an outstanding tool if SCO is applicable to spin dynamics simulations.
The basic idea of SCO is switching long-range interactions $V_{l}$ stochastically either to zero or a pseudo-interaction $\overline{V}_l$ with keeping the detailed balance condition, where $l$ denotes the bond index.
When we implement SCO into spin dynamics simulations, each time evolution of spin motion is obviously different from one without using SCO because the bond update process changes effective field from surrounding spins.
However, it is nontrivial how the time evolution of probability distribution for spins is modified by the implementation.
Since the probability distribution is generally described by Fokker-Planck (FP) equation, it is essential to compare the FP equation with and without using SCO.
More practically, when the coefficients of FP equation, diffusion and drift coefficients, are constructed to be the same, we can conclude that the both methods give the same spin dynamics.

To simulate spin dynamics, the s-LLG equation is generally adopted.
However, it is difficult to derive the FP equation with implementing SCO into s-LLG due to the unclear problem how to handle the bond update process of SCO in terms of stochastic differential equation.
For this problem, we adopt time quantified Monte Carlo (TQMC) method~\cite{Nowak2000a,Chubykalo2003,Cheng2005,Cheng2006}.
It has been proved that TQMC gives the same FP equation as s-LLG by tuning the time quantification factors~\cite{Cheng2005,Cheng2006}.
Since TQMC is based on the MC simulations, the above problem does not arise, and we can derive FP equation for TQMC with SCO (TQMC+SCO).
As a result, we find that the FP equation of TQMC+SCO is rigorously equivalent to that of TQMC.

To demonstrate TQMC+SCO, we simulate a magnetization reversal process in a simple cubic lattice system.
With this comparison, we find that the time evolution of probability distribution for total magnetization coincides with that calculated by the s-LLG equation.
This fact indicates that TQMC+SCO is an outstanding tool of spin dynamics simulations for long-range interacting systems with complicated lattice structures.

In addition, we propose several technical improvements of the method of TQMC.
These improvements enable us to take larger value of time step with the same precision comparing to the original TQMC~\cite{Nowak2000a,Chubykalo2003,Cheng2005,Cheng2006}.

The present paper is organized as follows:
In Sec.~\ref{sec:scoeq}, SCO for Monte Carlo simulation is briefly reviewed.
In Sec.~\ref{sec:scods}, we introduce TQMC+SCO and analytically prove that TQMC+SCO gives the same dynamics of the s-LLG.
In Sec.~\ref{sec:scomod}, we reconsider the equations in SCO into suitable forms for the real-time dynamics simulation.
In Sec.~\ref{sec:numc}, we demonstrate the method, and confirm the validity by comparing the results to those of the s-LLG.
In Sec.~\ref{sec:discussion}, summary and discussion are given.
The TQMC is briefly reviewed in Appendix~\ref{sec:appA}.
The technical improvements of TQMC are explained in Appendix~\ref{sec:appB}.

\section{SCO Method\label{sec:scoeq}}

First, we briefly review SCO for equilibrium state calculations.
Let us consider, as an example, a classical Heisenberg spin system $\left\{ \bm{s}_1,\bm{s}_2,\ldots, \bm{s}_N\right\}$ with long-range interactions $\{V_l = V_l\left(\bm{s}_{l_1},\bm{s}_{l_2}\right)\}$, where $l$ denotes the bond index.
Instead of using the original interaction $V_l$, SCO stochastically rejects the $l$-th bond interaction with probability $p_l$ or accept the pseudo-interaction $\overline{V}_l$ with the probability $1-p_l$.
According to Ref.~\onlinecite{Mak2005a}, $\overline{V}_l$ and $p_l$ can be determined so as to keep the detailed balance condition as follows:
\begin{align}
	\overline{V}_{l} = V_{l}- \beta^{-1}\ln\left[1 - p_l\right],
	\label{eq:pseudo_p}
\end{align}
and
\begin{align}
	p_l = \exp\left[\beta\left(V_l - V_l^\ast\right)\right],
	\label{eq:probability}
\end{align}
respectively.
Here, $V_l^{\ast}$ is a constant greater than or equal to the maximum value of $V_l$, and $\beta$ denotes the inverse temperature.
The rejection probability $p_l$ approaches to 1 when the maximum value of $V_l$ approaches to zero.
Therefore, long distant bonds are almost rejected by performing the bond update process, and thus we can significantly reduce the computational time of spin update process.

On the other hand, the bond update is performed over all the long-range interacting bonds, and thus naively takes a $\mathcal{O}\left(N^2\right)$ computational time.
Previous studies, however, proposed efficient algorithms that can compute this process in $\mathcal{O}(\beta N\ln N)$~\cite{Sasaki2008,Hinokihara2018}.
Among them, the algorithm proposed in Ref.~\onlinecite{Hinokihara2018} does not take advantage of the translational symmetries of systems.
Therefore, SCO is applicable even for complicated lattice systems.

\section{spin dynamics simulations with SCO \label{sec:scods}}

Next, let us discuss the idea of SCO is also applicable to the spin dynamics simulations.
Namely, implementing SCO does not change the time evolution of probability distribution, which is given by the FP equation.
For a classical Heisenberg spin system in the spherical coordinates, the general form of the FP equation is given by 
\begin{align}
	\frac{d}{dt}P\left(\left\{\theta\right\},\left\{\phi\right\},t\right)
	=& -\sum_i \frac{\partial}{\partial\theta_i}\left(A_{\theta_i}P\right)
	-\sum_i \frac{\partial}{\partial\phi_i}\left(A_{\phi_i}P\right)\nonumber\\
	&+\frac{1}{2}\sum_{i,j} \frac{\partial^2}{\partial\theta_i\partial\theta_j}\left(B_{\theta_i\theta_j}P\right)\nonumber\\
	&+\frac{1}{2}\sum_{i,j} \frac{\partial^2}{\partial\theta_i\partial\phi_j}\left(B_{\theta_i\phi_j}+B_{\theta_j\phi_i}P\right)\nonumber\\
	&+\frac{1}{2}\sum_{i,j} \frac{\partial^2}{\partial\phi_i\partial\phi_j}\left(B_{\phi_i\phi_j}P\right).
	\label{eq:general_FPE}
\end{align}
Here, $P\left(\left\{\theta\right\},\left\{\phi\right\},t\right)$, $A$, and $B$ denote the probability distribution, the drift coefficients, and the diffusion coefficients, respectively.
$A$ and $B$ are defined as the ensemble mean of an infinitesimal change of $\theta$ and $\phi$ with respect to time, e.g., $A_{\theta} = \mathrm{lim}_{\Delta t\rightarrow 0} \langle \Delta \theta\rangle/\Delta t$, and $B_{\theta\phi} = \mathrm{lim}_{\Delta t\rightarrow 0} \langle \Delta \theta \Delta \phi\rangle/\Delta t$.
The FP equation is characterized by these FP coefficients.

The previous study for TQMC calculated the FP coefficients for the spin update process consisting of a random spin motion and a precessional motion~\cite{Cheng2005}.
As an example, the drift term for TQMC $A_{\theta_i}^{\mathrm{TQMC}}$ is given by
\begin{widetext}
	\begin{align}
		\label{eqn:FP_Adrift}
		A_{\theta_i}^{\mathrm{TQMC}} &= \lim_{\Delta t \to 0} \frac{1}{\Delta t} \left\{\frac{\left[r^3\right]}{8}\beta\left(-\pdv{E}{\theta_i}  +  \frac{1}{\beta} \cot\theta_i \right) - \frac{\Phi}{\sin\theta_i}\pdv{E}{\phi_i} + \mathcal{O}\qty(\left[r^4\right])\right\}.
	\end{align}
\end{widetext}
Here, $E$ and $\beta$ is a total energy of the system and the inverse temperature, respectively.
$[r^n]$ and $\Phi$ are quantities appearing in the procedure of TQMC (see Appendix~\ref{sec:appA}).
By tuning parameters in TQMC, it is derived that all the FP coefficients are the same as those of s-LLG.
We review the detail of TQMC in Appendix~\ref{sec:appA} and suggest the technical improvement for the method of TQMC in Appendix~\ref{sec:appB}.

Now, we study how the FP coefficients are modified by implementing SCO.
At a glance, the idea of SCO seems to be incompatible with the real-time dynamics with the following reasons.
The advantage of SCO is a significant reduction of long-range interacting bonds with keeping the detailed balance condition.
In the spin dynamics simulation, however, the significant reduction of bonds changes the time evolution of magnetization trajectory.
In addition, although the representation of $\overline{V}_l$ and $p_l$ are derived so as to keep the detailed-balance condition, 
keeping the condition does not ensure to provide the correct real-time dynamics.

\begin{figure*}[tp]
	\centering
	\includegraphics[width=\hsize,keepaspectratio]{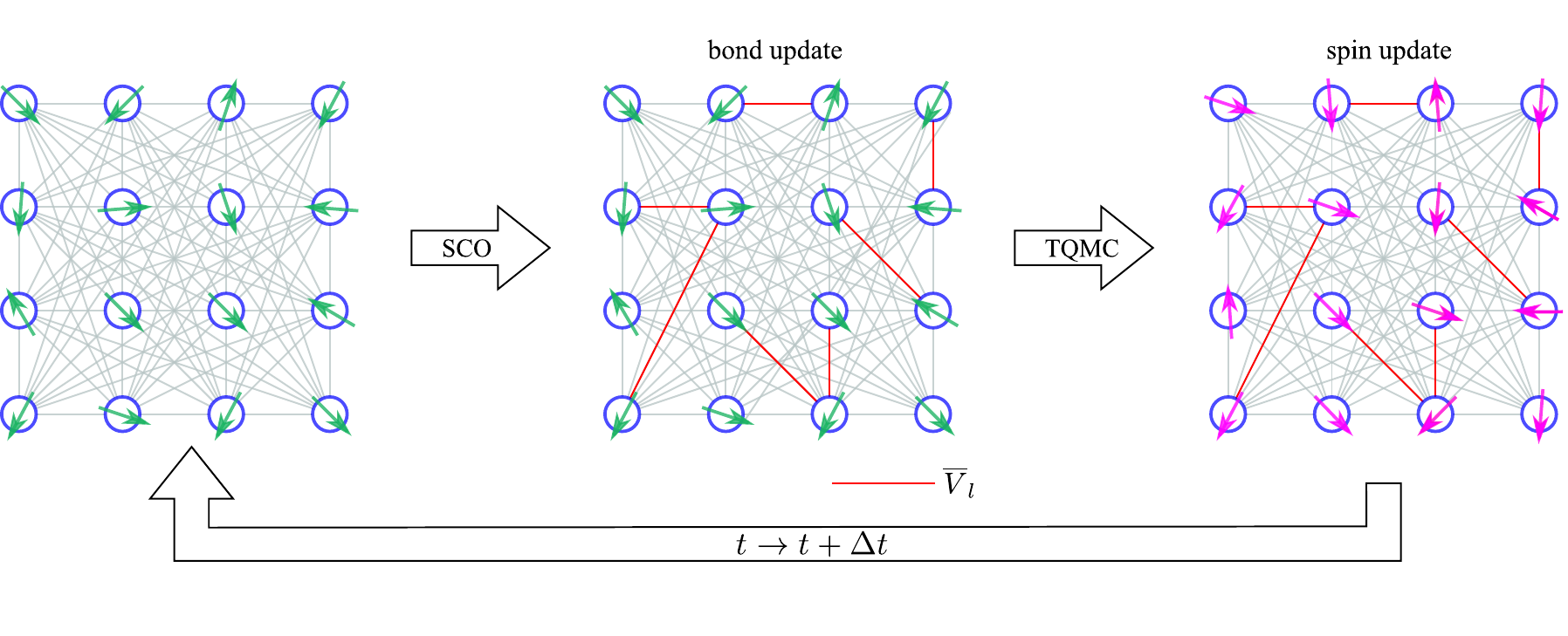}
	\caption{Schematic picture of the time proceeding process of TQMC+SCO.}
\label{fig00}
\end{figure*}

Nevertheless, we found that SCO does not change all the FP coefficients.
TQMC+SCO proposed in this paper consists of the following procedures.
\begin{description}
	\item[(i)] Perform the bond update process using SCO with a spin configuration at time $t$.
	\item[(ii)] Perform the spin update process using TQMC.
	\item[(iii)] Proceed time to $t+\Delta t$
	\item[(iv)] Return to step (i)
\end{description}
The schematic picture of this method is shown in Fig.~\ref{fig00}.
The bond update process is the same as SCO as we mentioned in the previous section.

Introducing SCO modifies the derivation of FP coefficients as follows.
First, SCO modifies the total energy $E$ to $E_{\mathrm{SCO}}$ defined as
\begin{align}
	E_{\mathrm{SCO}} &= E_{\mathrm{short}} + \sum_{l} \delta_{g_l,1}\overline{V}_{l}\nonumber\\
	&= E_{\mathrm{short}} + \sum_{l} \delta_{g_l,1}\left(V_l - \beta\ln\left[1-p_l\right] \right),
	\label{eq:e_sco}
\end{align}
where $E_{\mathrm{short}}$ denotes the energy of the exchange couplings, the external field, and the anisotropy.
$\delta_{g_l,1}$ denotes the Kronecker delta, and $g_l$ represents the bond state, which is accepted ($g_l=1$) or rejected ($g_l=0$).

Second, since TQMC+SCO randomly selects a bond configuration $\left\{g_l\right\}$, taking the average of bond configurations is required to calculate the FP coefficients.
The expectation value for an arbitrary function $X$, which depends on a bond configuration, is calculated as follows: 
\begin{align}
	\langle X\left(\left\{g_n\right\}\right)\rangle_{\mathrm{SCO}} = \sum_{\left\{ g_n\right\}} \prod_{n} \left(\qty(1-p_n)\delta_{g_n,1} +p_n\delta_{g_n,0}\right)  X\left(\left\{g_n\right\}\right).
	\label{eq:ex_sco}
\end{align}
Considering the above two modifications, Eqs.~\eqref{eq:e_sco} and \eqref{eq:ex_sco}, the drift term for TQMC+SCO $A_{\theta_i}^{\mathrm{SCO}}$ can be calculated as
\begin{widetext}
	\begin{align}
		\label{eq:Asco}
		A_{\theta_i}^{\mathrm{SCO}} &= \lim_{\Delta t \to 0} \frac{1}{\Delta t} \left\{\frac{\left[r^3\right]}{8}\beta\left(-\expval{\pdv{E_\mathrm{SCO}}{\theta_i}}_\mathrm{SCO}  +  \frac{1}{\beta} \cot\theta_i \right) - \frac{\Phi}{\sin\theta_i}\expval{\pdv{E_\mathrm{SCO}}{\phi_i}}_\mathrm{SCO} + \mathcal{O}\qty(\left[r^4\right])\right\}.
	\end{align}
\end{widetext}
% Note that SCO method only changes the total energy and does not affect the expectation values of $r^n$.
Substituting Eqs.~\eqref{eq:pseudo_p}, \eqref{eq:probability}, and \eqref{eq:e_sco} into the derivatives of $E_{\mathrm{SCO}}$ in terms of $\theta_i$ and $\phi_i$,
the derivative in Eq.~\eqref{eq:Asco} is given by
\begin{align}
	\Big\langle\pdv{E_{\mathrm{SCO}}}{\chi}\Big\rangle_{\mathrm{SCO}}=\pdv{E_{\mathrm{short}}}{\chi} + \Big\langle\sum_{l} \delta_{g_l,1}\frac{1}{1-p_l}\pdv{V_l}{\chi}\Big\rangle_{\mathrm{SCO}},
	\label{eq:desco}
\end{align}
where $\chi$ denotes $\theta_i$ or $\phi_i$.
By using Eq.~\eqref{eq:ex_sco}, the second term in the rhs is given by 
\begin{align}
	\Big\langle &\sum_{l} \delta_{g_l,1}\frac{1}{1-p_l}\pdv{V_l}{\chi}\Big\rangle_{\mathrm{SCO}}\nonumber\\
	&=\sum_{\left\{g_n\right\}}\prod_{n}\left(\qty(1-p_n)\delta_{g_n,1} + p_n\delta_{g_n,0}\right)\sum_l\delta_{g_l,1}\frac{1}{1-p_l}\pdv{V_l}{\chi}\nonumber\\
	&=\sum_{l}\prod_{n\neq l} \left\{\sum_{g_n=0,1}\left(\qty(1-p_n)\delta_{g_n,1} + p_n\delta_{g_n,0}\right)\right\}\nonumber\\
	&\times \left\{\sum_{g_l=0,1}\left(\qty(1-p_l)\delta_{g_l,1} + p_n\delta_{g_l,0}\right)\delta_{g_l,1}\frac{1}{1-p_l}\pdv{V_l}{\chi}\right\}\nonumber\\
	&=\sum_{l}\pdv{V_l}{\chi}.
	\label{eq:res_desco2}
\end{align}
This result ensures the following relation:
\begin{align}
	\Big\langle\pdv{E_{\mathrm{SCO}}}{\chi}\Big\rangle_{\mathrm{SCO}}=\pdv{E_{\mathrm{short}}}{\chi} + \sum_{l}\pdv{V_l}{\chi} = \pdv{E}{\chi}.
	\label{eq:desco_equivalence}
\end{align}
Therefore, both TQMC and TQMC+SCO give the same drift term: $A_{\theta_i}^{\rm SCO}=A_{\theta_i}^{\rm TQMC}$.

It should be noted that the other FP coefficients (see Eqs.~\eqref{eq:appFP2},~\eqref{eq:appFP3},~\eqref{eq:appFP4}, and~\eqref{eq:appFP5}) contain square of the first derivatives of total energy.
These terms cause a difference between TQMC and TQMC+SCO.
However, the influence is negligible at zero time step limit $\Delta t \rightarrow 0$ (see Eqs.~\eqref{eqn:relation1} and \eqref{eqn:relation2}).
Therefore, all the FP coefficients are rigorously equivalent to those for TQMC at zero time step limit.
This indicates a surprising result that the reduction of bonds in accordance with SCO does not change the real-time dynamics.

\section{Arbitrariness of parameters in SCO \label{sec:scomod}}

In the previous section, we proved that TQMC+SCO gives the same real-time dynamics of TQMC.
Note that we employed the expressions of Eqs.~\eqref{eq:pseudo_p} and \eqref{eq:probability}, which are derived so as to keep the detailed balance condition.
Meanwhile, it is found that the inverse temperature dependence of Eqs.~\eqref{eq:pseudo_p} and \eqref{eq:probability} vanishes after taking the average of bond configurations as seen in Eq.~\eqref{eq:desco_equivalence}.
To make the use of this fact, in TQMC+SCO, we can modify the expression of $\overline{V}_l$ and $p_l$ as follows:
\begin{align}
	\overline{V}_{l} = V_{l}- \tilde{\beta}^{-1}\ln\left[1 - p_l\right],
	\label{eq:pseudo_pt}
\end{align}
and
\begin{align}
	p_l = \exp\left[\tilde{\beta}\left(V_l - V_l^\ast\right)\right].
	\label{eq:probabilityt}
\end{align}
Here, $\tilde{\beta}$ is a positive parameter, which can be different from the inverse temperature $\beta$ included in Eq.~\eqref{eqn:FP_Adrift}.
We can easily check that the above representation also gives the same FP coefficients as TQMC.

The number of accepted bonds decreases as $\tilde{\beta}$ decreases.
Thus, we can reduce a computational cost by taking a smaller value of $\tilde{\beta}$.
However, in practical numerical calculation, taking a smaller value of $\tilde{\beta}$ makes the spin dynamics different due to the finite time step $\Delta t$.
We will discuss the relation between the values of $\tilde{\beta}$ and $\Delta t$ in the following section.

\section{Numerical simulation using TQMC+SCO\label{sec:numc}}

In this section, we demonstrate TQMC+SCO with a finite time step applying to the magnetization reversal process.
We study the classical-spin system whose Hamiltonian is given by
\begin{align}
	\mathcal{H} &= -\sum_{\langle i,j\rangle } J \bm{s}_i\cdot\bm{s}_j - \sum_{i} K\left(\bm{s}_i\cdot \bm{e}_z \right)^2 \nonumber\\
	&+\sum_{l} V_l\left(\bm{s}_{l_1},\bm{s}_{l_2}\right) - \sum_{i} \bm{H}\cdot\bm{s}_i,
	\label{hamiltonian}
\end{align}
where
\begin{align}
	V_l\left(\bm{s}_{l_1},\bm{s}_{l_2}\right)  = D\left(\frac{\bm{s}_{l_1}\cdot\bm{s}_{l_2}}{r_{l_1l_2}^3} - 3 \frac{\bm{s}_{l_1}\cdot{\bm{r}_{l_1l_2}}\bm{s}_{l_2}\cdot{\bm{r}_{l_1l_2}}}{r_{l_1l_2}^5}\right),
\end{align}
on a $10\times10\times10\times$ cubic lattice with open boundary conditions.
Here, $J$, $K$, $H$, and $D$ denote the nearest neighbor exchange coupling, the anisotropy, the external field, and the amplitude of the DDI, respectively. 
$\bm{r}_{l_1l_2}$ denotes the vector for the distance between $l_1$ and $l_2$ sites.
In the following, we simulate the time evolution of the magnetization reversal process by applying the external field oriented at $\pi/4$ from the easy axis.
We adopt $D=0.05J$, $T=0.2J$, $H=0.085J$, and $K=0.1J$.
We consider the magnetization reversal process given by the s-LLG dynamics with the damping constant $\alpha$.
The time step $\Delta t$ and the damping factor $\alpha$ take the values of $0.0001$ and $0.01$, respectively.
We take $V^\ast_{l} = \max\left[ V_l\left(\bm{s}_{l_1},\bm{s}_{l_2}\right)\right]$ and $\tilde{\beta}= \beta$.

\begin{figure}[tp]
	\centering
	\includegraphics[width=\hsize,keepaspectratio]{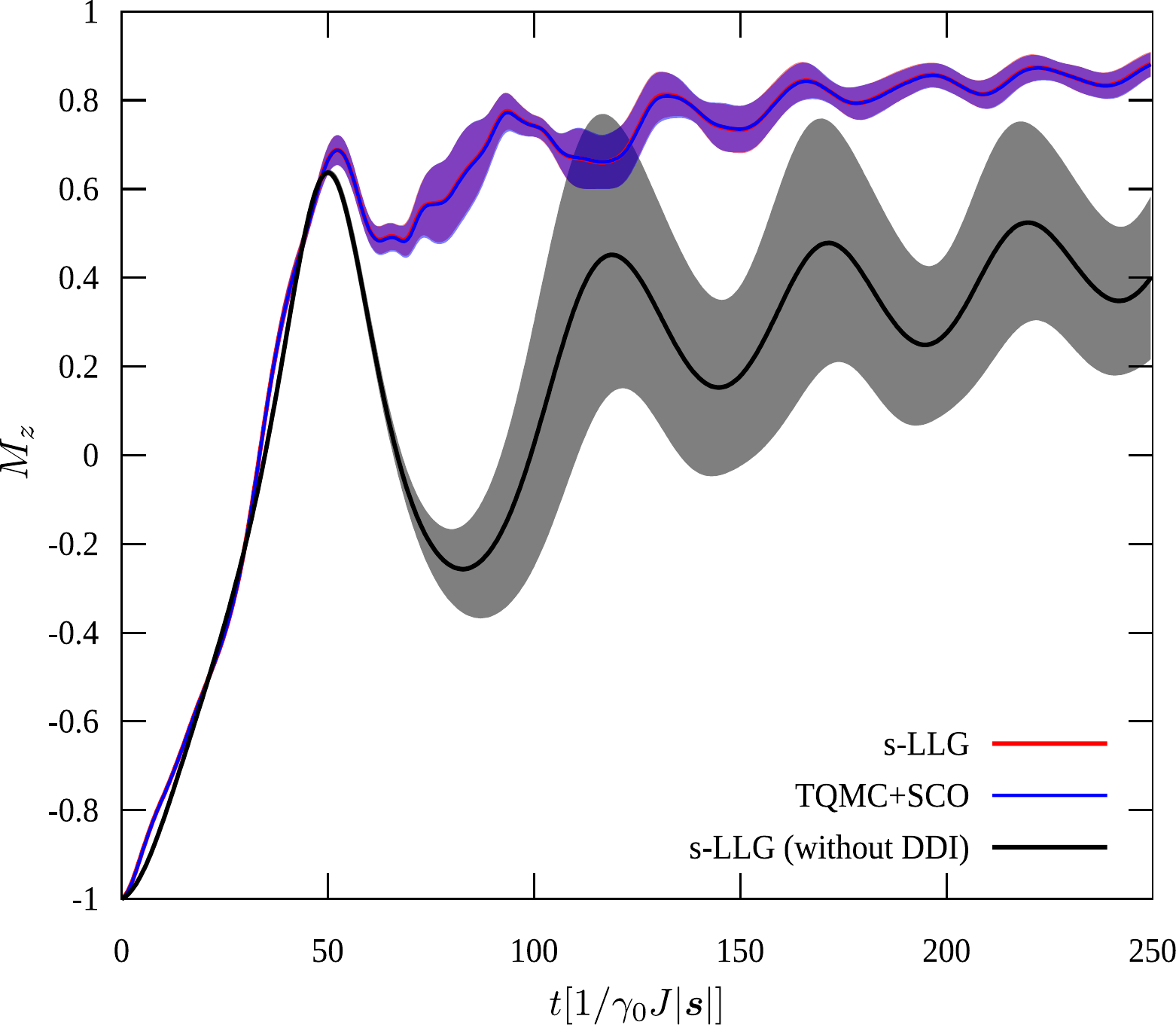}
	\caption{Magnetization reversal processes: the s-LLG with DDI (red line), TQMC+SCO (blue line), and the s-LLG without DDI (black line). The shade area denotes the standard deviation $\sigma$ of each method.}
	\label{fig1}
\end{figure}

In Fig.~\ref{fig1}, we depict the average (line) and the interval of the standard deviation $\sigma$ (shaded area) for 3000 samples.
The magnetization trajectories of systems with and without DDI are considerably different.
Thus, the effect of DDI is significant in the present case.
By using SCO, we confirmed that the number of accepted bonds per site is of the order of $10$, which indicates that most bonds are rejected.
Nevertheless, Fig.~\ref{fig1} indicates that TQMC+SCO gives the same real-time dynamics as the s-LLG method.

\begin{figure}[tp]
	\centering
	\includegraphics[width=\hsize,keepaspectratio]{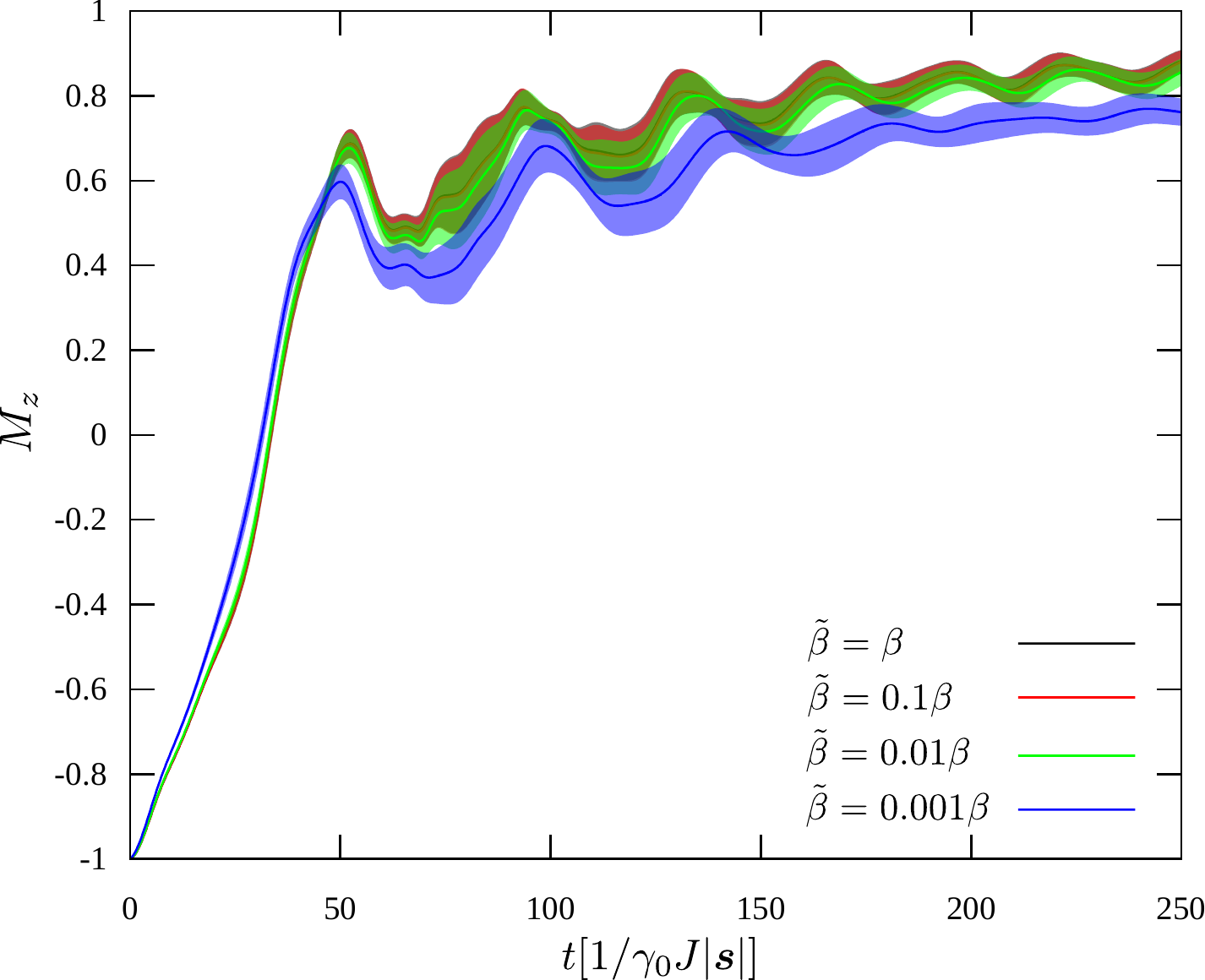}
	\caption{ Magnetization reversal processes with different values of $\tilde{\beta}$: $\tilde{\beta}=\beta$ (black line), $\tilde{\beta}=0.1\beta$ (red line), $\tilde{\beta}=0.01\beta$ (green line), and $\tilde{\beta}=0.001\beta$ (blue line).}
	\label{fig2}
\end{figure}
\begin{figure}[tp]
	\centering
	\includegraphics[width=\hsize,keepaspectratio]{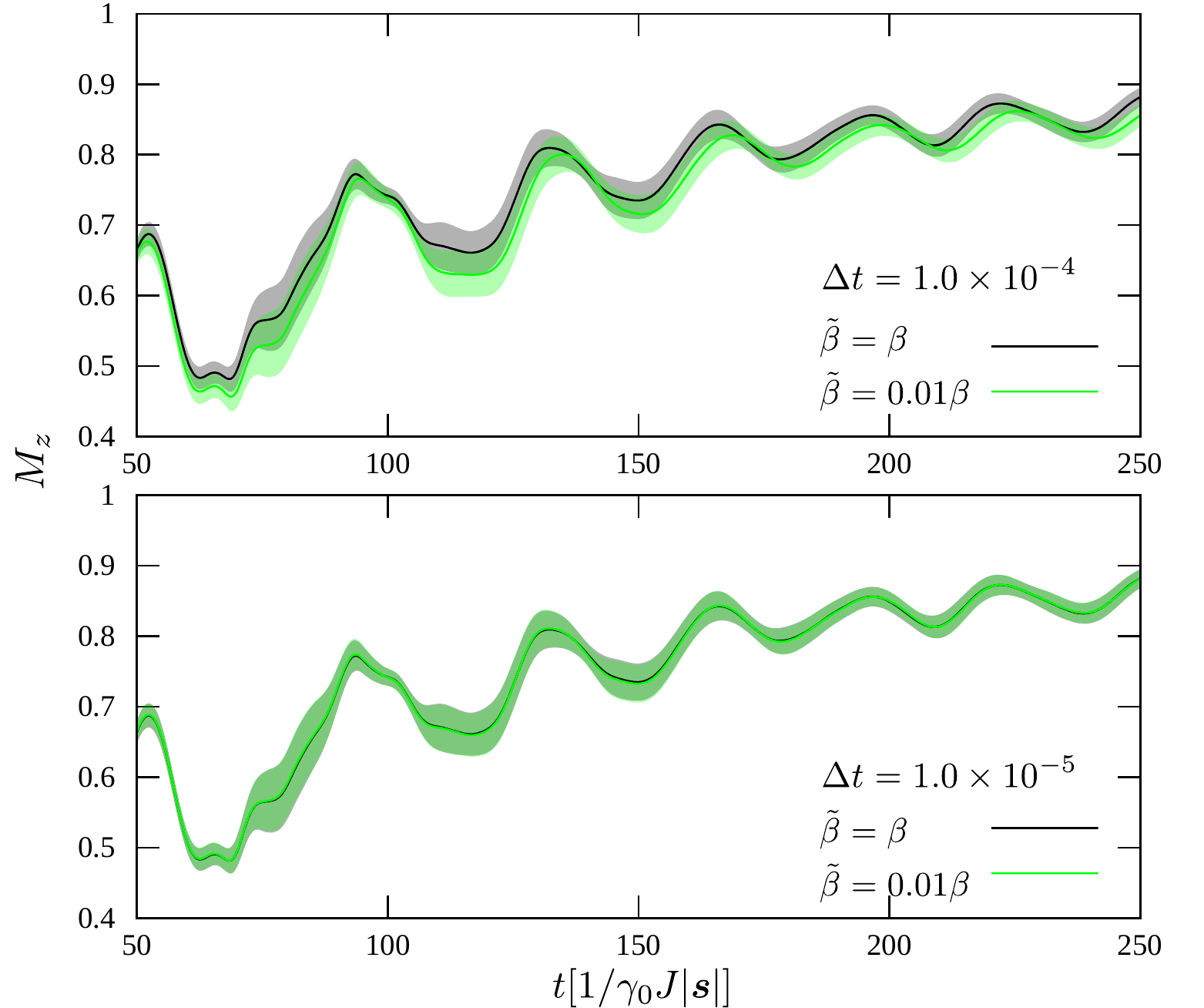}
	\caption{ Time step dependence of the Magnetization reversal trajectory for $\tilde{\beta}=\beta$ (black line) and $\tilde{\beta}=0.01\beta$ (green line). Time step is taken as $\Delta t = 1.0\times 10^{-4}$ (top panel) and $\Delta t = 1.0\times 10^{-5}$ (bottom panel).}
	\label{fig3}
\end{figure}

As we mentioned in the previous section, the bond update process does not depend on $\tilde{\beta}$ at least in the zero time step limit.
In the numerical simulation, however, depending on the value of $\Delta t$, the magnetization trajectory may change due to higher order terms for $\Delta t$ including $\tilde{\beta}$.
To study this effect, we simulate cases with $\tilde{\beta}=\beta$, $\tilde{\beta}=0.1\beta$, $\tilde{\beta}=0.01\beta$, and $\tilde{\beta}=0.001\beta$ as shown in Fig.~\ref{fig2}, where other parameters are the same as the case of Fig.~\ref{fig1}.
The magnetization trajectory for $\tilde{\beta}=0.1\beta$ is almost same as that for $\tilde{\beta}=\beta$.
Although difference appears in other cases ($\tilde{\beta}=0.01\beta$ and $\tilde{\beta}=0.001\beta$), we confirmed that the difference becomes small by decreasing the time step as shown in Fig.~\ref{fig3}.
These results indicate that TQMC+SCO is surely independent with the value of $\tilde{\beta}$ for $\Delta t\rightarrow 0$ limit.
% Thus, $\Delta t$ and $\tilde{\beta}$ have a trade-off relation.

\begin{figure}[tp]
	\centering
	\includegraphics[width=\hsize,keepaspectratio]{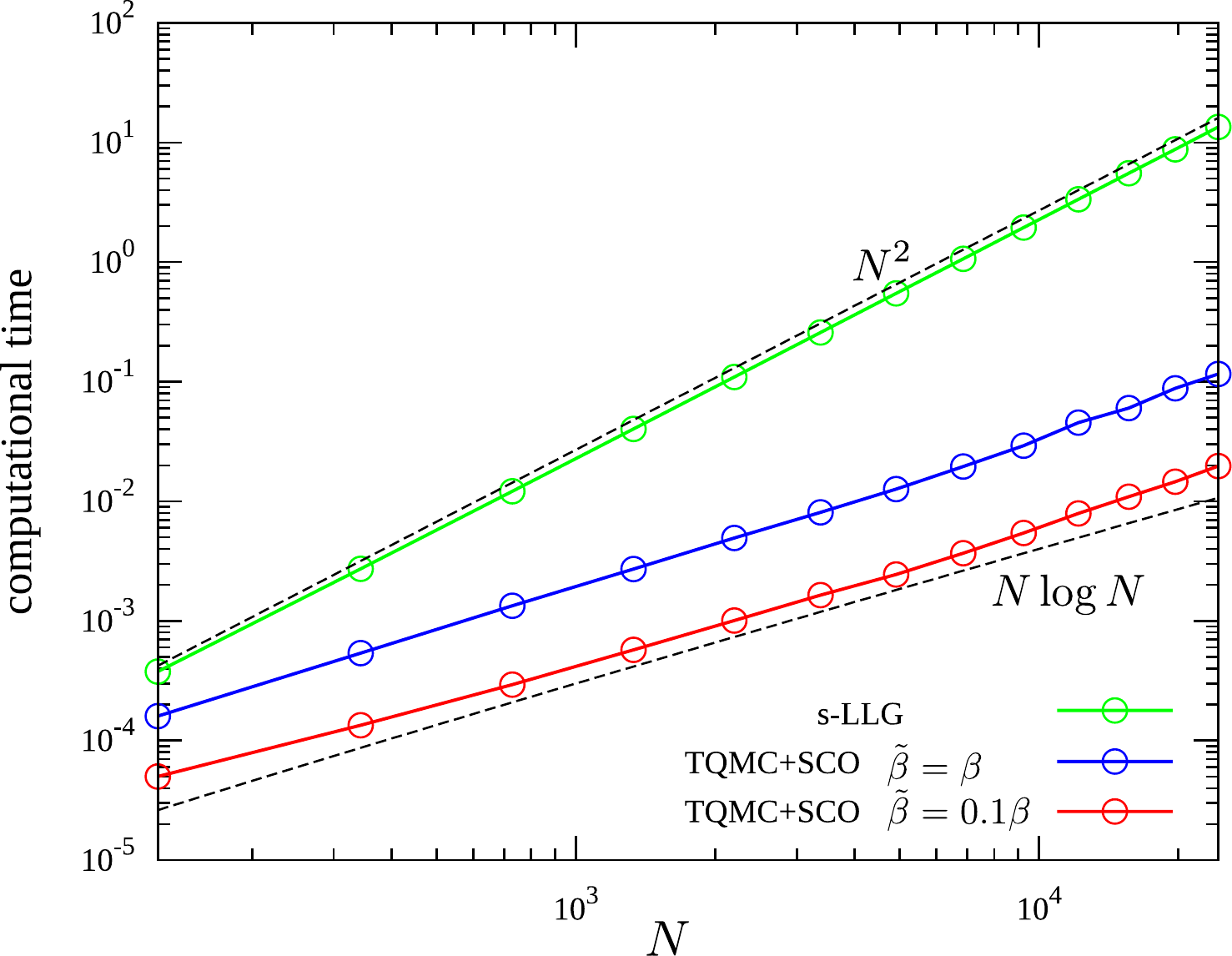}
	\caption{Computational time by the s-LLG (green line), TQMC+SCO with $\tilde{\beta}=\beta$ (blue line), and TQMC+SCO with $\tilde{\beta}=0.1\beta$ (red line).  $N$ denotes the number of spins in the system.}
	\label{fig4}
\end{figure}

The system size dependence of computational time of TQMC+SCO and that of the s-LLG equation are shown in Fig.~\ref{fig4}.
Here, in the s-LLG equation, the DDI is calculated without using the FFT, and thus the computational time is proportional to $N^2$, where $N$ denotes the number of spins.
On the other hand, TQMC+SCO is roughly proportional to $\tilde{\beta}N\log N$ as pointed out in Ref.~\onlinecite{Sasaki2008}.
As shown in Fig.~\ref{fig4}, the computational time is reduced by decreasing $\tilde{\beta}$.
Here, as we see in Fig.~\ref{fig3}, the value of $\Delta t$ was taken enough small to reproduce the correct real-time dynamics for $\tilde{\beta}=\beta$ and $\tilde{\beta}=0.1\beta$.
However, the precision depends on $\tilde{\beta}$, i.e., the deviation of magnetization trajectory becomes large as $\tilde{\beta}$ decreases.
To suppress the deviation, we need to take smaller $\Delta t$, which increases the computational time.
Hence, a kind of trade-off relation exists between the value of $\tilde{\beta}$ and $\Delta t$.
Optimization of these values are required for efficient numerical simulations.

\section{Summary and Discussion\label{sec:discussion}}

We developed TQMC+SCO and proved the validity.
At zero time step limit, we analytically proved that TQMC+SCO gives the same real-time dynamics as the s-LLG.
Since SCO reduces the number of bonds, the magnetization trajectory for each sample is obviously deviated from the s-LLG.
However, we found that the time evolution of probability distribution still gives the correct results even if we introduce SCO.
In the finite time step cases, we confirmed that results calculated by TQMC+SCO are in good agreement with those calculated by the s-LLG method by demonstrating the magnetization reversal process.

We also confirmed that TQMC+SCO reduces the computational time significantly, i.e. $O\left(\tilde{\beta} N\log N \right)$ for three dimensional systems.
Although the computational time presented in this paper is calculated without parallelization, TQMC+SCO can, in principle, be parallelized without any difficulty.
Moreover, since SCO algorithm does not require the translational properties of the system, we can apply TQMC+SCO to complicated lattice systems, which cannot be calculated by using the FFT.
Therefore, TQMC+SCO is a new and powerful method to study real-time dynamics of long-range interacting systems.

\acknowledgments
We would like to thank Takashi Mori for useful discussions and information.
This work is supported by the Elements Strategy Initiative Center for Magnetic Materials (ESICMM) under the outsourcing project of MEXT.
The authors thank the Supercomputer Center, the Institute for Solid State Physics, The University of Tokyo, for the use of the facilities.

\appendix
\section{ FP coefficients for TQMC\label{sec:appA}}

In this appendix, we briefly review TQMC~\cite{Cheng2005,Cheng2006}.
The essence of TQMC is to design spin update procedure to give the same FP coefficients as those of the s-LLG equation.
Hereafter, we consider the classical Heisenberg spin system.
In the spherical coordinates, the general form of the FP equation is given by Eq.~\eqref{eq:general_FPE}.

The s-LLG equation for the $N$ spin system is written as
\begin{align}
	\frac{d}{dt} \left\{\bm{s}\right\} = \frac{\gamma_0}{1+\alpha^2}\left( \left\{\bm{s}\right\} \times \left(\left\{\bm{h}\right\} + \alpha \left\{\bm{s}\right\} \times \left\{\bm{h}\right\} \right)\right).
	\label{eq:LLG}
\end{align}
Here, $\alpha$ and $\gamma_0$ are the damping constant and the gyromagnetic ratio, respectively.
$\bm{h}_i$ denotes the magnetic field at $i$-th site.
Note that $\bm{h}_i$ consists of the thermal fluctuation $\bm{\xi}_i$ and the effective field $\bm{h}^\mathrm{eff}_i$ from the surrounding environments~\cite{Garcia-Palacios1998a,Nishino2015}:
\begin{align}
	\bm{h}_i &=  \bm{h}^\mathrm{eff}_i+ \bm{\xi}_i \nonumber\\
	&= - \nabla_i E+ \bm{\xi}^t_i,
	\label{eq:LLG_h}
\end{align}
where $E$ is the total energy of the system.
The thermal fluctuation $\bm{\xi}_i$ is the white Gaussian noise and the following properties are assumed:
\begin{align}
	\expval{\xi^{\mu}_i}  &= 0,\\
	\expval{\xi^{\mu}_i\left(t\right) \xi^{\nu}_j \left(t^\prime\right)} &= 2\frac{\alpha}{\gamma_0\beta}\delta{i,j}\delta_{\mu\nu} \delta\left(t - t^\prime\right),
\end{align}
where $\mu$ and $\nu$ represent $x,y$, and $z$ components; $\beta$ denotes the inverse temperature.
According to Ref.~\onlinecite{Brown1963}, the FP coefficients for the s-LLG equation are given by 
\begin{align}
	A_{\theta_i}^{\mathrm{LLG}} &= \frac{\gamma_0}{1+\alpha^2}\left(-\alpha\frac{\partial E}{\partial \theta_i} + \frac{\alpha}{\beta}\cot\theta_i- \frac{1}{\sin\theta_i}\frac{\partial E}{\partial\phi_i}\right),\\
	A_{\phi_i}^{\mathrm{LLG}}   &=  \frac{\gamma_0}{1+\alpha^2}\left(\frac{1}{\sin\theta_i} \frac{\partial E}{\partial \theta_i} - \frac{\alpha}{\sin^2\theta_i}\frac{\partial E}{\partial\phi_i}\right),\\
	B_{\theta_i\theta_j}^{\mathrm{LLG}} &=\frac{2\alpha\gamma_0}{\left(1+\alpha^2\right)\beta} \delta_{ij},\\
	B_{\phi_i\phi_j}^{\mathrm{LLG}} &=\frac{2\alpha\gamma_0}{\left(1+\alpha^2\right)\beta} \frac{1}{\sin^2\theta_i}\delta_{ij},\\
	B_{\theta_i\phi_j}^{\mathrm{LLG}} &= B_{\phi_i\theta_j}^{\mathrm{LLG}} = 0,
	\label{eq:FPC_LLG}
\end{align}

In order to reproduce the FP coefficients, TQMC consists of two kinds of spin update processes: the precession motion and the random motion.
As an example, the drift coefficient of TQMC $A_{\theta_i}^{\rm TQMC}$ is defined as
\begin{align}
	A_{\theta_i}^{\rm TQMC}&\equiv \lim_{\Delta t \to 0}\frac{1}{\Delta t}\langle\Delta \theta_{i}\rangle\nonumber\\
	&=\lim_{\Delta t \to 0}\frac{1}{\Delta t}\left(\Delta \theta^\mathrm{prec}_i +\langle\Delta\theta^\mathrm{rand}_i\rangle_{0}\right).
	\label{eqn:Drift_TQMC}
\end{align}
Here, $\Delta \theta^\mathrm{prec}_i$ is the change of angle due to the precession motion.
This value is deterministic and is given by
\begin{align}
	\Delta \theta^\mathrm{prec}_i &\simeq -\Phi \bm{e}_{\theta i} \cdot\left(\bm{s}_i\times\bm{h}^{\mathrm{eff}}_i\right) \nonumber\\ 
	&= -\frac{\Phi}{\sin\theta_i} \pdv{E}{\phi_i},
	\label{eqn:precession}
\end{align}
where $\Phi$ is a parameter that will be tuned later so as to give same result of Eq.~\eqref{eq:FPC_LLG}.
\begin{figure}[t!]
	\begin{center}
	\includegraphics[width=\hsize,keepaspectratio]{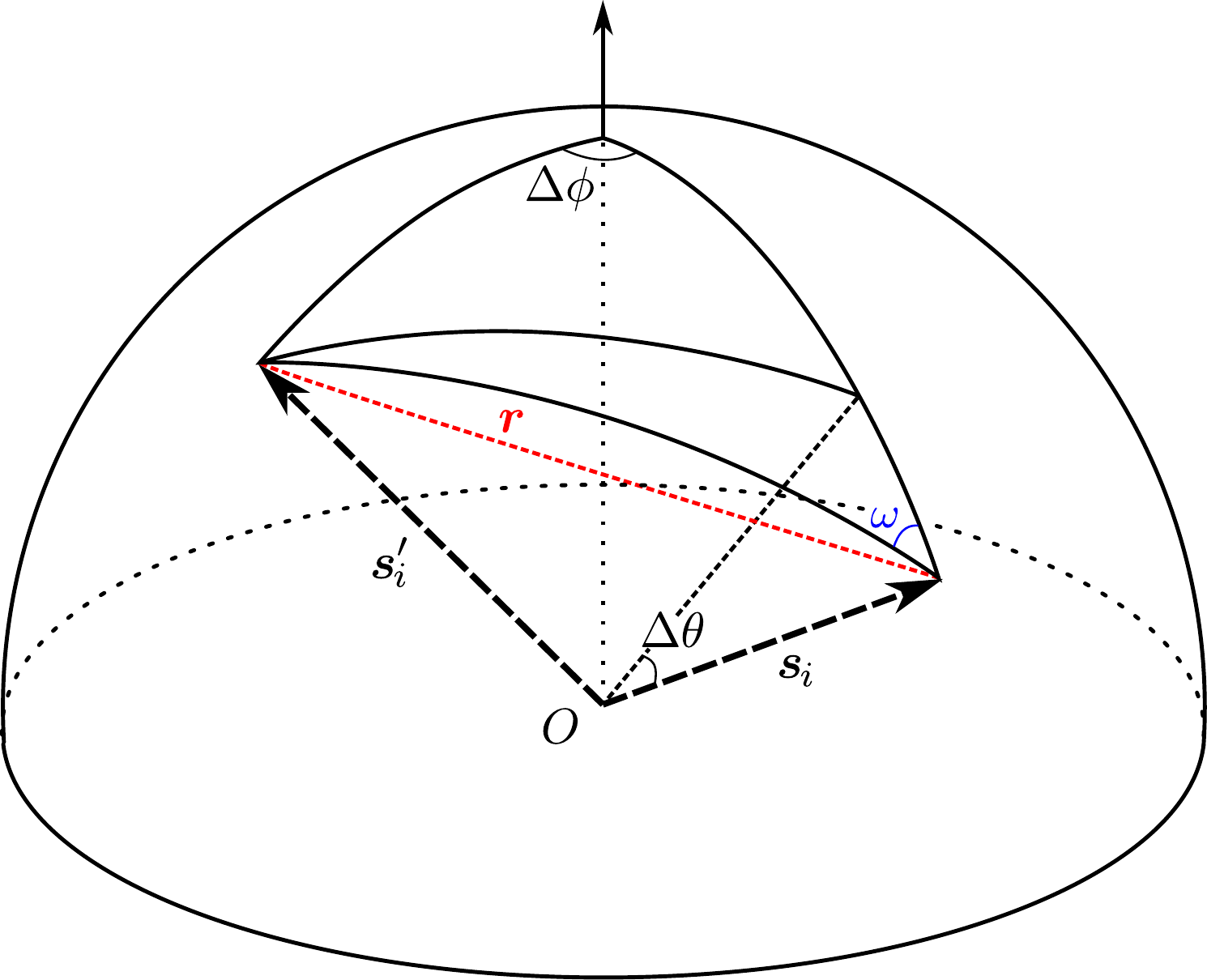}
	\end{center}
	\caption{Two random variables $r$ and $\omega$ which define angular changes caused by TQMC.}
	\label{fig:def_ralpha}
\end{figure}

On the other hand, the random spin motion from the spin $\bm{s}_i$ is generated as follows:
We first choose a candidate $\bm{s}^\prime_i$ and stochastically accept $\bm{s}^\prime_i$ in accordance with the heat-bath method.
Thus, the expectation value of the random spin motion is calculated as,
\begin{align}
	\langle\Delta\theta^\mathrm{rand}_i\rangle = \int d\bm{r}_i  P\qty(\bm{r}_i) A\qty(\Delta E_i) \Delta \theta^\mathrm{rand}_i.
	\label{eqn:TQMC0}
\end{align}
where $\bm{r}_i$ is the difference between $\bm{s}_i$ and $\bm{s}_i^\prime$,
$P\qty(\bm{r}_i)$ is the probability that $\bm{s}_i^\prime$ is chosen,
and $A\qty(\Delta E_i)$ is the acceptance probability of $\bm{s}_i^\prime$.
Since we assume that the length of $\bm{r}_i$ is small,  $A\qty(\Delta E_i)$ is approximately written as,
\begin{align}
	A\left(\Delta E_i\right)
	&=\frac{1}{1+\exp\qty(\beta\Delta E_i)}
	\nonumber\\
	&\approx  \frac{1}{2} - \frac{\beta}{4} \left( \pdv{E}{\theta_i}\Delta\theta^\mathrm{rand}_i + \pdv{E}{\phi_i}\Delta\phi^\mathrm{rand}_i\right).
	\label{eqn:taylor}
\end{align}

As shown in Ref.~\onlinecite{Cheng2006}, 
the change of $\theta^{\rm rand}_i$ and $\phi^{\rm rand}_i$ can be represented by using two random variables $r$ and $\omega$ as 
\begin{equation}
	\Delta \theta^{\rm rand}_i=-r\cos\omega+\frac{r^2}{2}\cot\theta\sin^2\omega+{\cal O}(r^3),
	\label{eqn:theta_rand}
\end{equation}
\begin{equation}
	\Delta \phi^{\rm rand}_i=r\frac{\sin\omega}{\sin\theta}+\frac{r^2}{2}\frac{\cot\theta}{\sin\theta}\sin 2\omega+{\cal O}(r^3), 
	\label{eqn:phi_rand}
\end{equation}
where $r$ denotes the amplitude of $\bm{r}$, and $\omega$ denotes the spherical surface angle measured from $-\bm{e}_\theta$ (see Fig.~\ref{fig:def_ralpha}). 

Let us assume that the probability $P\qty(\bm{r}_i)$ is isotropic on $\omega$.
Then, Eq.~\eqref{eqn:Drift_TQMC} can be calculated as follows:
\begin{widetext}
	\begin{align}
		A_{\theta_i}^{\mathrm{TQMC}} &= \lim_{\Delta t \to 0} \frac{1}{\Delta t} \left\{\frac{\left[r^3\right]}{8}\beta\left(-\pdv{E}{\theta_i}  +  \frac{1}{\beta} \cot\theta_i \right) - \frac{\Phi}{\sin\theta_i}\pdv{E}{\phi_i} + \mathcal{O}\qty(\left[r^4\right])\right\},
		\label{eq_App_A}
	\end{align}
where
\begin{align}
	\left[r^n\right] \equiv \int_0^\infty dr r^n P(r).
\end{align}
	Likewise, the other FP coefficients are given as 
	\begin{align}
	\label{eq:appFP2}
		A_{\phi_i}^{\mathrm{TQMC}} &= \lim_{\Delta t \to 0} \frac{1}{\Delta t} \left\{ - \frac{\left[r^3\right]}{8}\frac{1}{\sin^2\theta_i}\beta \pdv{E}{\phi_i}  + \frac{\Phi}{\sin\theta_i}\pdv{E}{\theta_i} + \mathcal{O}\qty(\left[r^4\right])\right\},\\
	\label{eq:appFP3}
		B_{\theta_i\theta_j}^{\mathrm{TQMC}} &= \lim_{\Delta t \to 0} \frac{1}{\Delta t} \left\{\frac{\left[r^3\right]}{4} +\left(\frac{\Phi}{\sin\theta_i}\pdv{E}{\phi_i}\right)^2 + \mathcal{O}\qty(\left[r^5\right])\right\}\delta_{i,j},\\
	\label{eq:appFP4}
		B_{\phi_i\phi_j}^{\mathrm{TQMC}} &= \lim_{\Delta t \to 0} \frac{1}{\Delta t} \left\{\frac{1}{\sin^2\theta}\frac{\left[r^3\right]}{4} +\left(\frac{\Phi}{\sin\theta_i}\pdv{E}{\theta_i}\right)^2 + \mathcal{O}\qty(\left[r^5\right])\right\}\delta_{i,j},\\
	\label{eq:appFP5}
		B_{\theta_i\phi_j}^{\mathrm{TQMC}} &= \lim_{\Delta t \to 0} \frac{1}{\Delta t} \left\{-2\frac{\Phi^2}{\sin^2\theta_i}\pdv{E}{\theta_i}\pdv{E}{\theta_j}  +\mathcal{O}\qty(\left[r^4\right])  \right\}\delta_{i,j}.
	\end{align}
\end{widetext}
By comparing these FP coefficients with those for the s-LLG, we find that the two sets of FP coefficients coincide with each other if the following relations are satisfied:
\begin{align}
	\label{eqn:relation1}
	\lim_{\Delta t\rightarrow 0}\left[ r^n\right]_0/\Delta t &= 
	\begin{cases}
		\displaystyle{8\frac{\alpha\gamma_0}{1+\alpha^2}} & n=3, \\
		\displaystyle{0} & n>3, \\
	\end{cases}\\
	\label{eqn:relation2}
	\Phi &= \frac{\gamma_0}{1+\alpha^2} \Delta t.
\end{align}
This fact indicates that we can make the FP coefficients for TQMC coincide with those for the s-LLG as long as $P\left(\bm{r}\right)$ is isotropic on $\omega$.
Note that $\left[r^n\right]$ depends on how we choose $\bm{r}_i$.

In the original TQMC, the computational procedure for the random spin motion is proposed as follows:
\begin{itemize}
	\item[(1a)] Pick a random vector lying within a sphere of radius $R$, where $R$ is a parameter 
		which controls the amplitude of the MC procedure.
	\item[(2a)] Add the vector generated in step (1) to $\bm{s}_i$ and normalize the resulting vector $\bm{s}^\prime_i$. 
	\item[(3a)] Accept the spin $\bm{s}^\prime_i$ generated in step (2) with the acceptance ratio of the heat-bath method $A(\Delta E_i)$
		, where $\Delta E_i$ is the energy difference caused by changing the spin from $\bm{s}_i$ to $\bm{s}^\prime_i$.
		Otherwise, the spin is unchanged.
\end{itemize}
Following the above procedure, the probability $P\left(\bm{r}\right)$ is isotropic on $\omega$ and given by
\begin{align}
	P(r)=
	\begin{cases}
		\displaystyle{\frac{3\sqrt{R^2-r^2}}{R^3}} & (0\le r \le R),\vspace{2mm}\\
		0 & (r>R).
	\end{cases}
	\label{eqn:def_P0}
\end{align}
Then, $\left[ r^n\right]$ is calculated as
\begin{align}
	\left[ r^n\right] = 
	\begin{cases}
		\displaystyle{\frac{2}{5}R^2} & n=3,\\
		\\
		\displaystyle{\frac{3\pi}{32}R^3} & n=4.\\
	\end{cases}
	\label{eqn:rn_tqmc}
\end{align}
Thus, comparing to Eq.~\eqref{eqn:relation1}, $R$ is determined as
\begin{align}
		R^2 = 20\frac{\alpha\gamma_0}{1+\alpha^2}\Delta t.
\end{align}
\section{improved TQMC\label{sec:appB}}

\begin{figure*}[t!]
	\centering
	\includegraphics[width=\hsize,keepaspectratio]{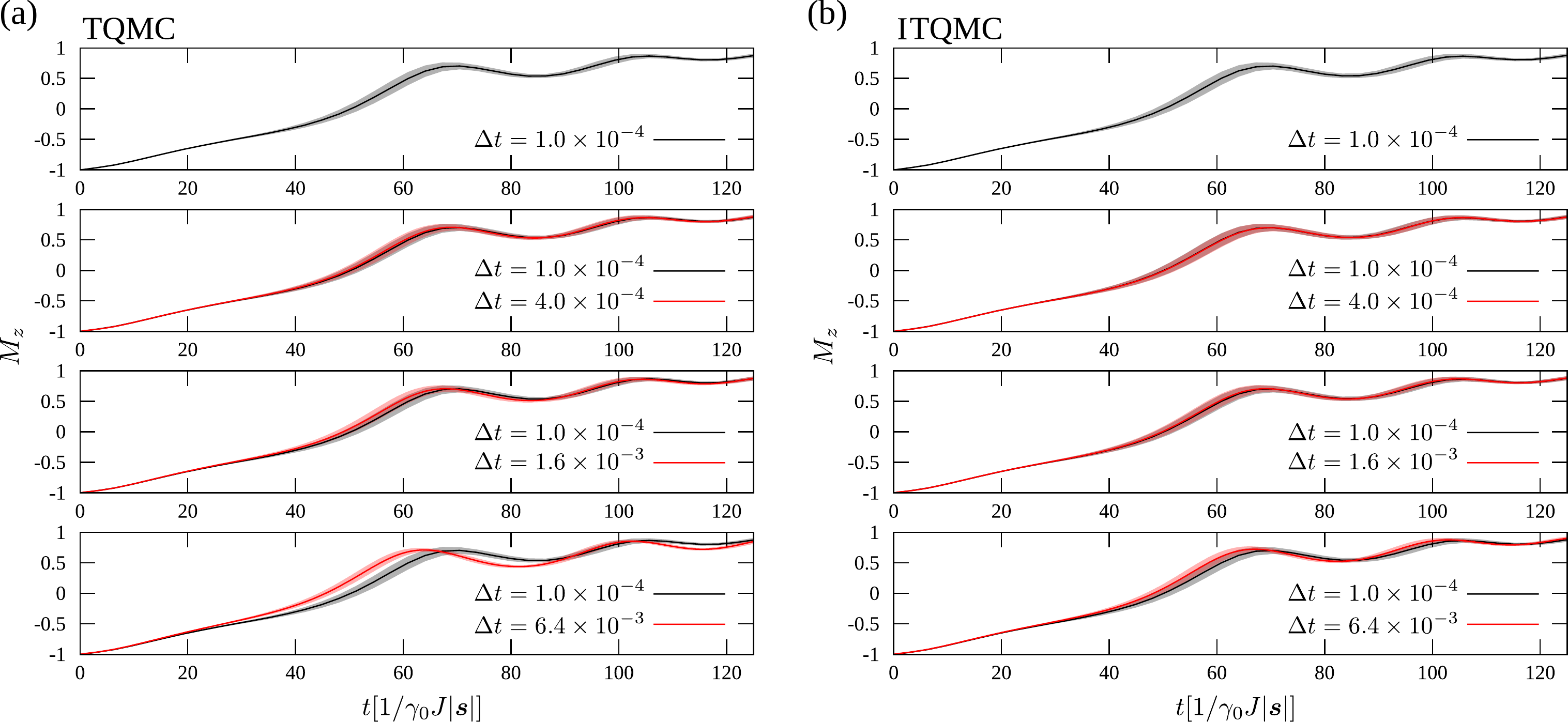}
	\caption{Magnetization reversal processes simulated by TQMC (a) and ITQMC (b) with different time steps: $\Delta t =1.0\times 10^{-4}$ (top panel), $\Delta t =4.0\times 10^{-4}$ (second panel), $\Delta t =1.6\times 10^{-3}$ (third panel), and $\Delta t =6.4\times 10^{-3}$ (bottom panel).
	The shaded area for each line indicates the standard deviation $\sigma$.
}
\label{fig0}
\end{figure*}
In the numerical simulation, we need to take a small but finite value of $\Delta t$.
Since taking large $\Delta t$ can reduce the computational time, it is important to consider how large $\Delta t$ we can take.
Thus, it is meaningful to evaluate the $\mathcal{O}\left(\sqrt{\Delta t}\right)$ terms, which are proportional to $[r^4]/\Delta t$, included in the FP coefficients before taking the zero time step limit.
This is because these higher order terms for $\Delta t$ makes FP coefficients different from that of s-LLG equation.
In the case of the original TQMC, $[r^4]/\Delta t$ is calculated as
\begin{align}
	[r^4]/\Delta t = \frac{3\pi}{32}R^3/\Delta t  =  \frac{3\pi}{32} \qty(\frac{40}{\beta}\frac{\alpha\gamma_0}{\left(1+\alpha^2\right)})^{3/2}\sqrt{\Delta t}.
\end{align}

Meanwhile, as mentioned above, we can make the FP coefficients for TQMC coincide with those for the s-LLG as long as $P\left(\bm{r}\right)$ is isotropic on $\omega$.
This fact indicates that it is possible to design the spin update procedure so as to reduce the deviation due to $[r^4]/\Delta t$.
Hereafter, we call this method improved TQMC (ITQMC).

The numerical procedure for the spin update process by using ITQMC is as follows:
\begin{itemize}
	\item[(1b)] Pick a random vector $\bm{r}_i$ lying on a {\it circle} of radius $R$.
	\item[(2b)] Add the vector $A\qty(\Delta E_i)\bm{r}_i$ to $\bm{s}_i$ and normalize to obtain the resulting vector $\bm{s}^\prime_i$. 
\end{itemize}
In step (2b), we employ the rejection free process by changing the adding vector from $\bm{r}_i$ to $A\qty(\Delta E_i)\bm{r}_i$.
Following the above procedure, the probability density  $P\left(\bm{r}_i\right)$ is also isotropic on $\omega$ and is expressed as
		\begin{align}
			P\qty(r)=  \frac{1}{R}\delta\qty(r-R).
			\label{eqn:defP1}
		\end{align}
Then, $\left[ r^n\right]$ is calculated as
\begin{align}
	\left[ r^n\right] =  R^{n-1}.
\end{align}
The FP coefficients for the ITQMC also coincide with those for the s-LLG if the following relation is satisfied:
\begin{align}
	\label{eqn:Irelation1}
	R^2 &= \frac{4\alpha\gamma_0}{\qty(1+\alpha^2)\beta}\Delta t.
\end{align}

Concluding this appendix, let us evaluate $\mathcal{O}\qty(\sqrt{\Delta t})$ terms that causes errors originating from the finiteness of the time step.
In common with TQMC, $\mathcal{O}\qty(\sqrt{\Delta t})$ terms are proportional to $[r^4]/\Delta t$, which is calculated as
\begin{align}
	[r^4]/\Delta t = R^3/\Delta t =  \qty(\frac{4\alpha\gamma_0}{\left(1+\alpha^2\right)\beta})^{3/2}\sqrt{\Delta t}.
\end{align}
Comparing to the original TQMC, the factor of $\sqrt{\Delta t}$ terms are suppressed as
\begin{align}
	\frac{16}{15\sqrt{10}\pi} \approx 0.107.
	\label{appB:ratio}
\end{align}
Therefore, it is expected that larger value of $\Delta t$ can be taken in ITQMC than TQMC.

Let us confirm the above expectation by demonstrating the magnetization reversal process.
We use the same spin lattice model as we employed in Sec.~\ref{sec:numc}.
Parameters are set as follows: $T=0.1J$, $H=0.082J$, $K=0.1J$, $D=0.0$, and $\alpha=0.1$.
Figure~\ref{fig0} shows the time step dependence of the magnetization reversal process calculated by TQMC and ITQMC.
We performed simulations for 1000 different samples with different random number sequences, and calculated the mean and standard deviation: solid line and shaded area denote the mean and interval of the standard deviation $\sigma$, respectively.
We confirmed that both methods give the same time evolution in the case of $\Delta t = 1.0\times 10^{-4}$.
By increasing the time step from this case, we evaluate how large time step we can take for both TQMC and ITQMC.

As seen in Fig.~\ref{fig0}, both methods indicate that the difference from the correct result ($\Delta t = 1.0\times 10^{-4}$) becomes larger as $\Delta t$ increases.
In the case of $\Delta t = 1.6\times 10^{-3}$, TQMC shows a deviation from the correct result, while ITQMC does not.
In the case of $\Delta t = 6.4\times 10^{-3}$, both methods give wrong results, but the deviations in ITQMC are smaller than those in TQMC.
Thus, we conclude that employing ITQMC enables us to take a larger value of $\Delta t$ than TQMC, and thus we can accelerate numerical simulations.

Finally, we also comment on the effect of the Gilbert damping $\alpha$.
The deviation due to the finite value of $\Delta t$ consists of a contribution of precession of the order of $\Phi^2$, and that of random motion given by $[r^4]$.
For small value of $\alpha$ the latter disappears, while the former remain finite.
Thus, difference between TQMC and ITQMC is not so large because ITQMC modifies only the random motion.
In contrast, for large value of $\alpha$ the deviation due to the finite value of $\Delta t$ mainly comes from $\mathcal{O}\left(\left[r^4\right]\right)$.
Thus, ITQMC has superiority for large value of $\alpha$ because the coefficient is largely reduced as Eq.~\eqref{appB:ratio}.

\bibliography{main3.bbl}
% \bibliography{lib.bib}
\end{document}